\documentclass[pra, showpacs]{revtex4}
\usepackage{amssymb}
\usepackage{bm}
\usepackage{graphicx}%

\begin{document}
\title
{Non-dissipative drag of superflow in a two-component Bose gas}

\author{D.\ V.\ Fil$^{1,2}$ and S.\ I.\ Shevchenko$^3$}

\affiliation{%
$^1$Institute for Single Crystals, National Academy of Sciences of
Ukraine, Lenin av. 60, Kharkov 61001, Ukraine\\ $^2$Ukrainian
State Academy of Railway Transport, Feyerbakh Sq. 7, 61050
Kharkov, Ukraine\\$^3$B.\, Verkin Institute for Low Temperature
Physics and Engineering, National Academy of Sciences of Ukraine,
Lenin av. 47 Kharkov 61103, Ukraine}

\begin{abstract}
A microscopic theory of a non-dissipative drag in a two-component
superfluid Bose gas is developed. The  expression for the drag
current in the system with the components of different atomic
masses, densities and scattering lengths is derived. It is shown
that the drag current is proportional to the square root of the
gas parameter. The temperature dependence of the drag current is
studied and it is shown that at temperature of order or smaller
than the interaction energy the temperature reduction of the drag
current is rather small. A possible way of measuring  the drag
factor is proposed. A toroidal system with the drag component
confined in two half-ring wells separated by two Josephson
barriers is considered. Under certain condition such  a system can
be treated as a Bose-Einstein counterpart of the Josephson charge
qubit in an external magnetic field. It is shown that the
measurement of the  difference of number of atoms in two wells
under a controlled evolution of the state of the qubit allows to
determine the drag factor.
\end{abstract}

% \submitto{\jpb}
\maketitle
\section{Introduction}
\label{intro}

Macroscopic quantum coherence manifests itself in many specific
phenomena. One of them is a non-dissipative drag that takes place
in superfluids and superconductors. The non-dissipative drag, also
known as the Andreev-Bashkin effect, was considered, for the first
time, in Ref. \cite{1}, where a three velocity hydrodynamic model
for $^3$He-$^4$He superfluid mixtures was developed. It was shown
that superfluid behavior of such systems can be described under
accounting the "drag"\ term in the free energy. This term is
proportional to the scalar product of the superfluid velocities of
two superfluid components. A similar situation may take place in
mixtures of superfluids of $S_z=+1$ and $S_z=-1$ pairs in liquid
$^3$He in the $A$-phase \cite{2}. Among other objects, where the
non-dissipative drag may be important, are neutron stars, where
the mixture of neutron and  proton Cooper pair Bose condensates is
believed to realize \cite{3,4}. The possibility of realization of
the non-dissipative drag in superconductors was considered in
\cite{5}. The non-dissipative drag in bilayer Bose systems was
treated microscopically in \cite{6,7} for a special case of two
equivalent layers of charged bosons. The case of a bilayer system
of neutral bosons was studied in \cite{my}  in the limit of small
interlayer interaction.

The most promising systems where the non-dissipative drag can be
observed experimentally are two-component alkali metal vapors. In
such systems the interaction between atoms of different species is
of the same order as the interaction between atoms of the same
specie and the effect is expected to be larger than in bilayers.
In Bose mixtures the components are characterized by different
densities, different masses of atoms and different interaction
parameters. In this paper we consider such a general case and
obtain an analytical expression for the drag current for zero and
finite temperatures.

In the system under consideration the drag force influences the
dynamics of atoms in the drag component  in the same manner  as
the vector potential of electromagnetic field influences  the
dynamics of electrons in superconductors. In particular, in
neutral superfluids with Josephson links the drag effect may
induce the gradient of the phase of the order parameter in the
bulk and, as a consequence, control the phase difference between
weakly coupled parts of the system. Therefore, on can expect that
the effect reveals itself in a modification of Josephson
oscillations between weakly coupled Bose gases. In this paper we
discuss possible ways for the observation such a modification. We
consider the Bose gas confined in a toroidal trap with two
Josephson links. In the Fock regime \cite{leg} the low energy
dynamics of the system can be described by the qubit model of
general form (the model, where all three components of the
pseudomagnetic field can be controlled independently). The
parameters of the qubit Hamiltonian depend on the drag factor. The
measurement of the state of the qubit under controlled evolution
allows to observe the effect caused by the non-dissipative drag
and determine the drag factor. In this paper we consider two
particular schemes of the measurement. In the first scheme one
should  determine the time required to transform a reproducible
initial state to a given final state. In the second scheme the
geometrical (Berry) phase should be detected.

In Sec. \ref{sec2} the microscopic theory of the non-dissipative
drag in two-component Bose gases is developed. In Sec. \ref{sec3}
a model of the Bose-Einstein qubit subjected by the drag force is
formulated and the schemes of measurement of the drag factor are
proposed. Conclusions are given in Sec. \ref{sec5}.

\section{Nondissipative drag in a two-component Bose system. Microscopic derivation}
\label{sec2}

Let  us consider a uniform two-component atomic Bose gas in a
Bose-Einstein condensed state. We will study the most general
situation where the densities of atoms in each component are
different from one another ($n_1\ne n_2$), the atoms of each
components have different masses ($m_1\ne m_2$) and the
interaction between atoms is described by three different
scattering lengths ($a_{11}\ne a_{22}\ne a_{12}$). The Hamiltonian
of the system can be presented in the form
\begin{equation}\label{1}
  H=\sum_{i=1,2} (E_i - \mu_i N_i)+\frac{1}{2}\sum_{i,i'=1,2}
  E_{ii'}^{int},
\end{equation}
where
\begin{equation}\label{2}
  E_i=\int d^3 r \frac{\hbar^2}{2 m_i}
  [\nabla\hat{\Psi}^+_i({\bf r})]\nabla\hat{\Psi}_i({\bf r})
 \end{equation}
is the kinetic energy,
\begin{equation}\label{2a}
  E_{i i'}^{int}=\int d^3  r
 \hat{\Psi}^+_i({\bf r})\hat{\Psi}^+_{i'}({\bf r})\gamma_{i i'}
 \hat{\Psi}_{i'}({\bf r})
 \hat{\Psi}_{i}({\bf r})
 \end{equation}
is the energy of interaction,   $\gamma_{ii}=4\pi\hbar^2
a_{ii}/m_i$ and $\gamma_{12}=2\pi\hbar^2(m_1+m_2)a_{12}/(m_1m_2)$
are the interaction parameters,  and $\mu_i$ are the chemical
potentials.

For the further analysis it is convenient to use the density and
phase operator approach (see, for instance, \cite{11,12}). The
approach is based on the following representation for the Bose
field operators
\begin{equation}\label{3}
  \hat{\Psi}_i({\bf r})=\exp\left[i\varphi_{i}({\bf r})+i\hat{\varphi}_i({\bf r})\right]
  \sqrt{n_i+\hat{n}_i({\bf r})},
\end{equation}
\begin{equation}\label{4}
  \hat{\Psi}_i^+({\bf r})=\sqrt{n_i+\hat{n}_i({\bf r})}
  \exp\left[-i\varphi_{i}({\bf r})-i\hat{\varphi}_i({\bf
  r})\right],
\end{equation}
where $\hat{n}_i$ and $\hat{\varphi}_i$ are the density and phase
fluctuation operators, $\varphi_{i}({\bf r})$ are the $c$-number
terms of the phase operators, which are connected with the
superfluid velocities  by the relation ${\bf
v}_i=\hbar\nabla\varphi_i/m_i$. In what follows we specify the
case of the superfluid velocities independent of ${\bf r}$.

Substituting Eqs.  (\ref{3}), (\ref{4}) into Eq. (\ref{1}) and
expanding it in series in powers of $\hat{n}_i$ and
$\nabla\hat{\varphi}_i$ we present the Hamiltonian of the system
in the following form
\begin{equation}\label{5}
  H=H_{0}+H_{2}+\ldots
\end{equation}
In (\ref{5}) the term
\begin{equation}\label{5a}
  H_0=V\Bigg(\sum_{i=1,2} \left[\frac{1}{2} m_i n_i
  {\bf v}_i^2
  +\frac{\gamma_{ii}}{2}n_{i}^2-\mu_in_i\right] +\gamma_{12}n_1
  n_2\Bigg)
\end{equation}
does not contain the operator part.  Here $V$ is the volume of the
system. The minimization conditions for the Hamiltonian $H_0$
yield the equations
\begin{equation}\label{5d}
\frac{1}{2}m _i {\bf v}_i^2+\gamma_{ii} n_i+\gamma_{12} n_{3-i}-
  \mu_i=0\quad (i=1,2).
\end{equation}
Under the conditions (\ref{5d}) the terms, linear in the density
fluctuation operators, vanish in the Hamiltonian. Taking into
account the $\nabla(n_i\nabla\varphi_i({\bf r}))=0$, we find that
the terms, linear in the phase fluctuation operators, vanish in
the Hamiltonian as well.

The part of the Hamiltonian quadratic in $\nabla\hat{\varphi}_i$
and $\hat{n}_i$
 operators reads as
\begin{eqnarray}\label{5c}
   H_2=\int d {\bf r} \Bigg(\sum_i\Bigg\{\frac{\hbar^2}{2
  m_i}\Bigg[\frac{\Big(\nabla\hat{n}_i({\bf r})\Big)^2}{4 n_i}
  +n_i\Big(\nabla\hat{\varphi}_i({\bf r})\Big)^2\Bigg] \cr  +
  \frac{\hbar {\bf v}_i}{2}\Big(\hat{n}_i({\bf r})\nabla\hat{\varphi}_i({\bf r})
  +[\nabla\hat{\varphi}_i({\bf r})]\hat{n}_i({\bf r})\Big)\cr
  +
  \frac{i\hbar^2}{2m_i}\Big([\nabla\hat{n}_i({\bf r})]\nabla\hat{\varphi}_i({\bf r})-
  [\nabla\hat{\varphi}_i({\bf r})]\nabla\hat{n}_i({\bf r})\Big)
  +\frac{\gamma_{ii}}{2}\Big(\hat{n}_i({\bf r})\Big)^2\Bigg\}
   + \gamma_{12}\hat{n}_1({\bf r})\hat{n}_2({\bf r})
  \Bigg).
\end{eqnarray}

The quadratic part of the Hamiltonian determines the spectra of
the elementary excitations. Hereafter we will neglect the higher
order terms in the Hamiltonian (\ref{5}). These terms describe the
scattering of the quasiparticles and they can be omitted if the
temperature is much smaller than the temperature of Bose-Einstein
condensation.

Let us rewrite the quadratic part of the Hamiltonian in terms of
the operators of creation and annihilation of the elementary
excitations. As the first step, we use the substitution
\begin{equation}\label{6}
  \hat{n}_i({\bf r})=\sqrt{\frac{n_i}{V}}\sum_{\bf k}
  e^{i{\bf k r}}\sqrt{\frac{\epsilon_{i k}}{E_{i k}}}
  \left[ b_{i}({\bf k})+b^+_{i}(-{\bf
  k})\right],
\end{equation}
\begin{equation}\label{7}
  \hat{\varphi}_i({\bf r})=\frac{1}{2i}\sqrt{\frac{1}{n_i V}}\sum_{\bf k}
  e^{i{\bf k r}}\sqrt{\frac{E_{i k}}{\epsilon_{i k}}}
  \left[ b_{i}({\bf k})-b^+_i(-{\bf
  k})\right],
\end{equation}
where operators $b_i^+$, $b_i$ satisfy the Bose commutation
relations. Here $\epsilon_{i k}=\hbar^2 k^2/2 m_i$ is the spectrum
of free atoms, and
\begin{equation}\label{7a}
 E_{lk}=\sqrt{\epsilon_{i k}(\epsilon_{i k}+2\gamma_{ii} n_i)}
\end{equation}
is the spectrum of the elementary excitations at $\gamma_{12}=0$
and ${\bf v}_i=0$. The substitution (\ref{6}), (\ref{7})  reduces
the Hamiltonian (\ref{5c}) to the form quadratic in  $b_i^+$ and
$b_i$ operators:
\begin{eqnarray}\label{8}
  H_2=\sum_{i{\bf k}} \left[{\cal E}_{i}({\bf k}) \left(
b^+_{i}({\bf k})  b_{i}({\bf
  k})+\frac{1}{2}\right)-\frac{1}{2}\epsilon_{i k}\right]
 \cr +\sum_{\bf k} g_k\left[b^+_{1}({\bf k})  b_{2}({\bf k})+b_{1}({\bf k})
   b_{2}({\bf
  -k})+h.c.\right].
\end{eqnarray}
Here
\begin{equation}\label{401}
  {\cal E}_{i}({\bf k})=E_{i k}+\hbar  {\bf
k v}_i
\end{equation}
and
\begin{equation}\label{402}
  g_k=\gamma_{12}  \sqrt{\frac{\epsilon_{1 k}\epsilon_{2 k}n_1 n_2}{E_{1 k}E_{2
k}}}.
\end{equation}

The Hamiltonian (\ref{8}) contains non-diagonal in Bose creation
and annihilation operator terms and  it can be diagonalized using
the standard procedure of u-v transformation \cite{13}. The result
is
\begin{equation}\label{14}
 H_2= \sum_{\bf k}\Bigg[\sum_{\lambda=\alpha,\beta} {\cal E}_{\lambda}({\bf k})\Bigg(
   \beta_\lambda^+({\bf k})  \beta_\lambda({\bf
  k})+\frac{1}{2}\Bigg)-\frac{1}{2}\sum_{i=1,2}\epsilon_{i
  k}\Bigg],
\end{equation}
where $\beta_{\lambda}^+(k)$  and $\beta_{\lambda}(k)$ are the
operators of creation and annihilation of elementary excitations.

The energies ${\cal E}_{\lambda}({\bf k})$ satisfy the equation
\begin{equation}\label{n1}
  \det\left(\matrix{{\bf A}-{\cal E}{\bf I}&{\bf B}\cr{\bf B}&{\bf A}+{\cal
  E}{\bf I}}\right)=0,
\end{equation}
where
\begin{equation}\label{n2}
  {\bf A}=\left(\matrix{{\cal E}_1({\bf k})&0&g_k&0\cr
  0&{\cal E}_1(-{\bf k})&0&g_k\cr
  g_k&0&{\cal E}_2({\bf k})&0\cr
  0&g_k&0&{\cal E}_2(-{\bf k})}\right)
\end{equation}
\begin{equation}\label{n2a}
  {\bf B}=\left(\matrix{0&0&0&g_k\cr
  0&0&g_k&0\cr
  0&g_k&0&0\cr
  g_k&0&0&0}\right)
\end{equation}
and ${\bf I}$ is the identity matrix.
%Having the spectra of
%elementary  excitations  one can obtain the relation between the
%supercurrents in each component are their superfluid velocities.

The densities of superfluid currents in two components can be
obtained from the relation
\begin{equation}\label{101n}
  {\bf j}_i=\frac{1}{V}\frac{\partial  F}{\partial {\bf
  v}_i},
\end{equation}
where $F$ is the free energy of the system. Here the quantity
${\bf j}_i$ is defined as the density of the mass current.

The free energy of the system, described by the Hamiltonian
(\ref{5}), is given by the formula
\begin{equation}\label{201n}
 F=H_0 +
 \frac{1}{2}\sum_{{\bf k}}\left[\sum_{\lambda=\alpha,\beta}
  {\cal E}_{\lambda}( {\bf k})
   - \sum_{i=1,2}\epsilon_{i k}\right]
   + T \sum_{{\bf k}}\sum_{\lambda=\alpha,\beta}
   \ln\left[1-\exp\left(-\frac{{\cal
  E}_{\lambda}( {\bf k})}{T}\right) \right].
 \end{equation}
 The second term in (\ref{201n}) is
the energy of the zero-point fluctuations and the third term is
the standard temperature dependent part of the free energy for the
gas of noninteracting  elementary excitations.

We specify the case of small superfluid velocities (much smaller
than the critical ones). In this case the currents can be
approximated by the expressions linear in ${\bf v}_i$. To obtain
these expressions we will find the free energy as series in ${\bf
v}_i$, neglecting the terms higher than quadratic.

At  ${\bf v}_1={\bf v}_2=0$ the equation (\ref{n1}) is easily
solved and the spectra are found to be
\begin{equation}\label{n3}
  E_{\alpha(\beta) k}=
  \Bigg(\frac{E_{1k}^2+E_{2k}^2}{2}\pm\sqrt{\frac{(E_{1k}^2-E_{2k}^2)^2}{4}+
  4 \gamma_{12}^2 n_1 n_2 \epsilon_{1 k} \epsilon_{2
  k}}\Bigg)^{1/2}.
\end{equation}
As required in the procedure \cite{13}, we take positive valued
solutions of Eq. (\ref{n1}). The energies (\ref{n3}) should be
real valued quantities. This requirement yields the common
condition for the stability of the two-component system:
$\gamma_{12}^2\leq \gamma_{11}\gamma_{22}$. If this condition were
not fulfilled, spatial separation of two components (at positive
$\gamma_{12})$ or a collapse (at negative $\gamma_{12}$) would
take place.

At nonzero superfluid velocities we present the solutions  of Eq.
(\ref{n1}) as series in ${\bf v}_i$:
\begin{eqnarray}\label{n4}
 {\cal E}_{\alpha}({\bf k})=E_{\alpha k}+\frac{1}{2}\hbar {\bf k
v}_1 \left(1+\frac{E_{1k}^2-E_{2k}^2}{E_{\alpha k} ^2-E_{\beta
k}^2}\right) + \frac{1}{2}\hbar {\bf k v}_2\left(1-\frac{E_{1
k}^2-E_{2 k} ^2}{E_{\alpha k} ^2-E_{\beta k}^2}\right)\cr  +
\frac{2 \gamma_{12}^2 n_1 n_2 \epsilon_{1 k} \epsilon_{2
k}\left(3E_{\alpha k}^2+E_{\beta k}^2\right)}{E_{\alpha
k}\left(E_{\alpha k}^2-E_{\beta k}^2\right)^3}\hbar^2\left({\bf k
v}_1-{\bf k v}_2\right)^2,
\end{eqnarray}

\begin{eqnarray}\label{n5}
{\cal E}_{\beta}({\bf k})=E_{\beta k}+\frac{1}{2}\hbar {\bf k
v}_1\left(1-\frac{E_{1k}^2-E_{2k}^2}{E_{\alpha k} ^2-E_{\beta
k}^2}\right)+ \frac{1}{2}\hbar {\bf k v}_2\left(1+\frac{E_{1
k}^2-E_{2 k} ^2}{E_{\alpha k} ^2-E_{\beta k}^2}\right)\cr -
\frac{2 \gamma_{12}^2 n_1 n_2\epsilon_{1 k} \epsilon_{2
k}\left(E_{\alpha k}^2+3E_{\beta k}^2\right)}{E_{\beta
k}\left(E_{\alpha k}^2-E_{\beta k}^2\right)^3}\hbar^2\left({\bf k
v}_1-{\bf k v}_2\right)^2.
\end{eqnarray}
 Note that at ${\bf v}_1={\bf v}_2={\bf v}$ the spectra (\ref{n4}), (\ref{n5}) are
 reduced to common expressions for the energies of quasiparticles in a moving
 condensate:
 ${\cal E}_{\alpha(\beta)}({\bf k})=E_{\alpha(\beta) k}+\hbar {\bf k
v}$.

Using Eqs. (\ref{201n}),(\ref{n4}) and (\ref{n5}) we obtain the
following expression for the free energy
\begin{equation}\label{28a}
  F= F_0+  \frac{V}{2}\Bigg[(\rho_{1}-\rho_{n1}){\bf v}_1^2
  + (\rho_{2}-\rho_{n2}){\bf v}_2^2 -\rho_{\rm dr}
  ({\bf v}_1-{\bf v}_2)^2\Bigg],
\end{equation}
where $F_0$ does not depend on ${\bf v}_i$. In (\ref{28a})
$\rho_i=m_i n_i$ are the mass densities, the quantities
\begin{equation}\label{n6}
  \rho_{n1}=-\frac{ m_1}{3 V }\sum_{\bf
k}\epsilon_{1k}\Bigg[\frac{d N_{\alpha
  k}}{d E_{\alpha k}}+\frac{d N_{\beta
  k}}{d E_{\beta k}}+ \frac{E_{1 k}^2-E_{2 k} ^2}{E_{\alpha k}
^2-E_{\beta k}^2} \left(\frac{d N_{\alpha
  k}}{d E_{\alpha k}}-\frac{d N_{\beta
  k}}{d E_{\beta k}}\right)\Bigg],
\end{equation}
\begin{equation}\label{n6a}
  \rho_{n2}=-\frac{ m_2}{3 V}\sum_{\bf
k}\epsilon_{2k}\Bigg[\frac{d N_{\alpha
  k}}{d E_{\alpha k}}+\frac{d N_{\beta
  k}}{d E_{\beta k}}- \frac{E_{1 k}^2-E_{2 k} ^2}{E_{\alpha k}
^2-E_{\beta k}^2} \left(\frac{d N_{\alpha
  k}}{d E_{\alpha k}}-\frac{d N_{\beta
  k}}{d E_{\beta k}}\right)\Bigg]
\end{equation}
describe the thermal reduction of the superfluid densities, and
the quantity
\begin{eqnarray}\label{n7}
 \rho_{dr}=\frac{4}{3 V }\sqrt{m_1 m_2} \sum_{\bf
k}\frac{\gamma_{12}^2 n_1 n_2
  \left(\epsilon_{1 k}\epsilon_{2 k}\right)^{3/2}}
  {E_{\alpha k} E_{\beta k}} \Bigg[\frac{1+N_{\alpha k} +N_{\beta
  k}}{(E_{\alpha k}+E_{\beta k})^3}- \frac{N_{\alpha k} -N_{\beta
  k}}{(E_{\alpha k}-E_{\beta k})^3}\cr+\frac{2 E_{\alpha k} E_{\beta
  k}}{\left(E_{\alpha k}^2-E_{\beta k}^2\right)^2}\left(\frac{d N_{\alpha
  k}}{d E_{\alpha k}}+\frac{d N_{\beta
  k}}{d E_{\beta k}}\right)\Bigg],
\end{eqnarray}
which we call the "drag density," yields the value of
redistribution of the superfluid densities between the components.
In Eqs. (\ref{n6})-(\ref{n7}) $N_{\alpha(\beta) k} =[\exp(
E_{\alpha(\beta) k}/T)-1]^{-1}$ is the Bose distribution function.

Using Eqs. (\ref{101n}), (\ref{28a}) we arrive to the following
expressions for the supercurrents
\begin{equation}\label{22}
  {\bf j}_1=(\rho_{1}-\rho_{n1}-\rho_{dr}){\bf v}_1
  +\rho_{{\rm dr}}{\bf v}_2,
\end{equation}
\begin{equation}\label{23}
  {\bf j}_2=(\rho_{2}-\rho_{n2}-\rho_{dr}){\bf v}_2
  +\rho_{{\rm dr}}{\bf v}_1.
\end{equation}
One can see that  at nonzero $\rho_{dr}$ the current of one
component contains the term proportional to the superfluid
velocity of the other component. It means that there is a transfer
of motion between the components. In particular, at $v_1=0$ the
current in the component 1 (${\bf j}_1=\rho_{dr} {\bf v}_2$) is
purely the drag current. Since $\rho_{dr}$ is the function of
$\gamma_{12}^2$ (see Eqs. (\ref{n7}) and (\ref{n3})) the drag
current does not depend on the sign of the interaction between the
components.

Eq. (\ref{n7}) is the main result of the paper.  This equation
yields the value of the drag for the general case of two-component
Bose system with components of different densities, different
masses of atoms, different interaction parameters, and for zero as
well as for nonzero temperatures. Moreover, this equation is valid
not only for the point interaction between the atoms, but for any
central force interaction. In the latter case the interaction
parameters $\gamma_{ik}$ in Eq. (\ref{n7}) and in the spectra
(\ref{n3}), (\ref{7a}) should be replaced with the Fourier
components of the corresponding interaction potentials.

To estimate the absolute value of the drag we, for simplicity,
specify the case $m_1=m_2=m$, that is realized when two components
are two hyperfine states of the same atoms.

At  $T=0$ Eq. (\ref{n7}) is reduced to
\begin{equation}\label{n11}
  \rho_{dr}=\frac{4 m}{3}  \int_0^\infty d \epsilon
\frac{\gamma_{12}^2 n_1  n_2 \nu(\epsilon)
\epsilon^{1/2}}{\sqrt{(\epsilon+w_1)
  (\epsilon+w_2)}\left(\sqrt{\epsilon+w_1}+\sqrt{\epsilon+w_2}\right)^3},
\end{equation}
where
 $$\nu(\epsilon)= \frac{m^{3/2}}{\sqrt{2}\pi^2\hbar^3}
  \sqrt{\epsilon}$$
  is the density of states for free atoms, and
$$ w_{1(2)}=\gamma_{11}n_1+\gamma_{22}n_2\pm\sqrt{(\gamma_{11}
n_1-\gamma_{22} n_2)^2+ 4\gamma_{12}^2 n_1 n_2 }. $$

 The integral in (\ref{n11}) can be
evaluated analytically. To present the answer in a compact form it
is convenient to introduce the dimensionless parameters $$
\eta=\frac{a_{12}^2}{a_{11} a_{22}}\quad \textrm{ and}\quad
\kappa=\sqrt{\frac{n_1 a_{11}}{n_2 a_{22}}}+\sqrt{\frac{n_2
a_{22}}{n_1 a_{11}}}.$$ ($0\leq\eta\leq 1$ and $\kappa\geq 2$)

Using these notations we have
\begin{equation}\label{n11b}
\rho_{dr}= \sqrt{\rho_1 \rho_2}\sqrt[4]{ n_1 a_{11}^3 n_2
a_{22}^3}\frac{ \eta}{\sqrt{\kappa}}F(\kappa,\eta),
\end{equation}
where
\begin{eqnarray}\label{n11c}
 F(\kappa,\eta)=\frac{256}{45\sqrt{2\pi}}
  \frac{(\kappa+3\sqrt{1-\eta})\sqrt{\kappa}}{
  \Bigg(\sqrt{\kappa+\sqrt{\kappa^2-4+4
  \eta}}+\sqrt{\kappa-\sqrt{\kappa^2-4+4
  \eta}}\Bigg)^{3}}.
\end{eqnarray}
Direct evaluation of Eq. (\ref{n11c}) shows that  at allowed
$\eta$ and $\kappa$ ($0\leq\eta\leq 1$ and $\kappa\geq 2$) the
factor $F(\kappa,\eta)$ is almost the constant (the range of
variation of $F$ is $[0.7\div 0.8]$) and one can neglect the
dependence of $F$ on the parameter of the system.

At $a_{11}n_1=a_{22}n_2$ we obtain from (\ref{n11b}) the following
approximate relation
\begin{equation}\label{1001}
  \rho_{dr}\approx
  \frac{1}{2}\rho_1\frac{a_{12}^2}{a_{11}a_{22}}\sqrt{n_1 a_{11}^3}=
   \frac{1}{2}\rho_2\frac{a_{12}^2}{a_{11}a_{22}}\sqrt{n_2 a_{22}^3}.
\end{equation}
If the density of one component is much larger than of the other
and $a_{11}\sim a_{22}$, the "drag density" is approximated as
\begin{eqnarray}\label{1002}
\rho_{dr}\approx 0.8 \rho_1\frac{a_{12}^2}{a_{22}^2}\sqrt{n_2
a_{22}^3}\quad \textrm{at}\quad n_1\ll n_2, \cr {\rho_{dr}}\approx
0.8 \rho_2\frac{a_{12}^2}{a_{11}^2}\sqrt{n_1 a_{11}^3}\quad
\textrm{at}\quad n_2\ll n_1.
\end{eqnarray}
One can see that the "drag density" is proportional to the square
root of the gas parameter. It means that the drag effect is larger
in  "less ideal" Bose gases.

\begin{figure}
%\begin{center}
\includegraphics[height=6cm]{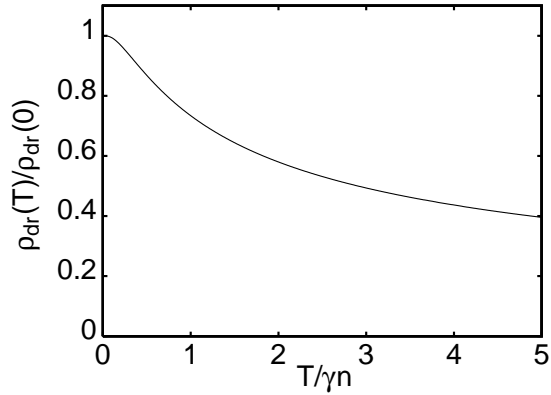}
\caption{\label{fig1}Dependence of the "drag density" on the
temperature.}
\end{figure}

The temperature dependence of the "drag density" at small $T$ can
be evaluated analytically from  Eq. (\ref{n7}) using the linear
approximation for the spectra of the excitations. It yields
$\rho_{dr}(T)=\rho_{dr}(0)(1-\alpha_T T^4/T_0^4)$, where
$T_0=\sqrt{\gamma_{11} n_1 \gamma_{22} n_2}$  and the factor
$\alpha_T$ is positive. Numerical evaluation of the sum over ${\bf
k}$ in Eq. (\ref{n7}) shows that the analytical approximation is
valid only at $T\ll T_0$. At $T \gtrsim T_0$ the "drag density"
decreases much slower under increase of the temperature. As an
example, the dependence of $\rho_{dr}(T)$  at $n_1=n_2=n$,
$\gamma_{11}=\gamma_{22}=\gamma$ and $\eta=0.5$ is shown in Fig.
\ref{fig1}.

Now let us discuss how the drag effect can reveal itself in a real
physical situation. If one deals with the stationary superflow one
implies that it is the circulating superflow, e.g., the tangential
superflow in a hole cylinder. In such a case the superfluid
velocities satisfy the Onsager-Feymnan quantization condition
\begin{equation}\label{d1}
  \oint {\bf v}_i d {\bf l} = \frac{2 \pi \hbar N_i}{m_i},
\end{equation}
where the vorticity parameters $N_i$ are integer. Then, the drag
effect can be understood as the appearance of the circulating
current in the drive component (e.g. specie 1), when the
circulation of the superfluid velocity of the drive component
(e.g. specie 2) is fixed ($N_2=const$). The current of the specie
1 (\ref{22}) depends on the superfluid velocities of the both
species and if the superfluid velocity of the drag component
directed anti-parallel to the superfluid velocity of the drive
component the current of the drag component might vanish. But
since the velocities are quantized it may happen only under
certain special conditions (see below). The superfluid velocity of
the drag component is determined by that at fixed $N_2$ the free
energy (\ref{28a}) has a minimum with respect to discrete values
of ${\bf v}_1=\hbar N_1/(m_1 R)$ (where the $R$ is the radius of
the contour in (\ref{d1})). Depending on the value of the
parameter
$\alpha=(\rho_{dr}/(\rho_2-\rho_{n2}-\rho_{dr}))(m_1/m_2)N_2$
several possibilities can be realized. At $|\alpha|<1/2$ the
minimum of the energy (\ref{28a}) corresponds to $N_1=0$ (and
${\bf v}_1=0$). In this case the current of the drag component is
directed along the drive current and it is proportional to the
drag density. At $|\alpha|=p$ ($p$ is natural) the value $N_1=-p$
minimizes the energy. In this case two terms in Eq. (\ref{22})
compensate each other and the current in the drag component
vanishes. At half-integer $\alpha$ the degenerate situation takes
place:  two state (with co-directed currents, and counter-directed
currents) have the same energy. At $1/2+p<|\alpha|<p$  the state
with counter-directed currents gains the energy and at
$p<|\alpha|<p+1/2$ the co-directed currents are energetically
preferable. In the latter two cases the nonzero vorticity of the
drag component ($N_1\ne 0$) is also induced. This behavior is
analogous the behavior of a superconducting ring in a magnetic
field. We note that since $\rho_{dr}\ll \rho_2$, the most
realistic case is $|\alpha|<1/2$ when the simple picture of the
transfer of part of the motion from the drive to the drag
component takes place.

In this study we have concentrated on the analytical derivation of
the drag effect in the uniform Bose gases. The  consideration of
the non-uniform case requires the solution of the eigenvalue
problem for the elementary excitations in the two-component Bose
gas in the external potential. But even for the simplest case of a
spherically symmetric trap this problem can be solved analytically
only in the long-wavelength limit and the Thomas-Fermi
approximation \cite{park} (the spectrum of elementary excitations
in one-component  Bose gases was obtained analytically for a
number of potentials but also in the same limit
\cite{12,14,15,16}). Since the main contribution to the drag
density comes from the excitations with the wave vectors of order
of the healing length (see (\ref{n7})), the rigorous analysis of
the drag effect can be done only numerically. Nevertheless, in the
Tomas-Fermi situation the drag effect can be evaluated basing on
the following arguments. When the linear size of the Bose cloud is
much larger than the healing length, the spectrum of the
excitations at the wave vectors of order or higher than the
inverse healing length is well described by the quasi uniform
approximation. Therefore, the drag effect can be described by the
same equations, as in the uniform case with the only modification
that the quantities $n_1$ and $n_2$, and, correspondingly,
$\rho_i$, $\rho_{ni}$, $\rho_{dr}$ and $j_i$ in Eqs.
(\ref{n6})-(\ref{23}) are understood as functions of coordinates.

At an arbitrary symmetry of the trap potential the superfluid
velocity of the drag component cannot be equal to zero in each
point. Indeed, in general case of space dependent $\rho_i$,
$\rho_{ni}$, and $\rho_{dr}$ the velocity field ${\bf v}_2({\bf
r})$ cannot satisfy two independent continuity conditions
$\nabla[(\rho_{2}-\rho_{n2}-\rho_{dr}){\bf v}_2]=0$ and
$\nabla(\rho_{dr} {\bf v}_2)$. To analyze this case one should
find the velocity fields ${\bf v}_1({\bf r})$ and ${\bf v}_2({\bf
r})$ that satisfy the continuity conditions and the quantization
conditions. To illustrate this point let us consider a simple
example of a trap having the shape of a hollow cylinder with the
densities that depend only on the polar angle $\phi$. We will seek
the velocity fields that do not have radial components. Then, the
Eqs. (\ref{22}), (\ref{23}), written for the tangential components
of the currents and the velocities, can be presented in the matrix
form
\begin{equation}\label{mm1}
  \left(\matrix{j_1\cr j_2}\right)=\hat{R}\left(\matrix{v_1(r,\phi)\cr
  v_2(r,\phi)}\right),
\end{equation}
where
\begin{equation}\label{mm2}
  \hat{R}=\left(\matrix{\rho_{s1}(\phi)-\rho_{dr}(\phi)&\rho_{dr}(\phi)\cr
  \rho_{dr}(\phi)&\rho_{s2}(\phi)-\rho_{dr}(\phi)}\right)
\end{equation}
with  $\rho_{si}(\phi)=\rho_i(\phi)-\rho_{ni}(\phi)$. Due to the
continuity conditions the current $j_1$ and $j_2$ in (\ref{mm1})
do not depend on $\phi$. According to Eq. (\ref{mm1}) the
velocities $v_1(r,\phi)$ and $v_2(r,\phi)$ are connected with the
currents by the equation
\begin{equation}\label{mm3}
\left(\matrix{v_1(r,\phi)\cr
  v_2(r,\phi)}\right)=\hat{R}^{-1}\left(\matrix{j_1(r)\cr j_2(r)}\right)
\end{equation}
Integrating Eq. (\ref{mm3}) over $\phi$ and taking into account
the quantization conditions (\ref{d1}) we obtain the equation for
the currents
\begin{equation}\label{mm4}
  \hat{T} \left(\matrix{j_1(r)\cr j_2(r)}\right)=\frac{2\pi\hbar}{r}
  \left(\matrix{N_1/m_1\cr N_2/m_2}\right),
\end{equation}
where
\begin{equation}\label{mm5}
  \hat{T}=\left(\matrix{\int_0^{2 \pi}d \phi \frac{\rho_{s2}-\rho_{dr}}
  {\rho_{s1}\rho_{s2}-\rho_{dr}(\rho_{s1}+\rho_{s2})}
  & -\int_0^{2 \pi}d \phi \frac{\rho_{dr}}
  {\rho_{s1}\rho_{s2}-\rho_{dr}(\rho_{s1}+\rho_{s2})}\cr
-\int_0^{2 \pi}d \phi \frac{\rho_{dr}}
  {\rho_{s1}\rho_{s2}-\rho_{dr}(\rho_{s1}+\rho_{s2})}&
\int_0^{2 \pi}d \phi \frac{\rho_{s1}-\rho_{dr}}
  {\rho_{s1}\rho_{s2}-\rho_{dr}(\rho_{s1}+\rho_{s2})}}
 \right)
\end{equation}
If a given vorticity of the drive component $N_2$ is not very
large the minimum of energy is reached at $N_1=0$. In the latter
case the solution  of Eq. (\ref{mm4}) in the leading order in
$\rho_{dr}$ yields the following expression for the current of the
drag component
\begin{equation}\label{ddd}
  j_1(r)\approx\frac{2\pi \hbar N_2}{m_2 r}\frac{\int_0^{2\pi}d \phi
  \frac{\rho_{dr}(\phi)}{\rho_{s1}(\phi)\rho_{s2}(\phi)}}
  {\int_0^{2\pi}d \phi \frac{1}{\rho_{s1}(\phi)}\int_0^{2\pi}d \phi \frac{1}{\rho_{s2}(\phi)}}
\end{equation}
One can see that if at some $\phi$ the density $\rho_{s1}$ has a
sharp minimum the first factor in denominator in Eq. (\ref{ddd})
becomes large. On the other hand, the integral in numerator is not
very sensitive to lowering of $\rho_{s1}$ (see Eqs. (\ref{1002})).
Thus, in a system with a "bottle neck" in the drag component the
drag current decreases strongly and the main consequence of the
drag effect is the emergence of the gradient of the phase of the
order parameter of the drag component. Similar situation takes
place in a system with a weak link. The latter case is analyzed in
the next section. In the uniform case Eq. (\ref{ddd}) is reduced
to $j_1=\rho_{dr} v_2$.

To complete the discussion we emphasize that the crossed  term
($\rho_{dr}{\bf v_1} {\bf v_2}$) in the free energy (\ref{28a})
(and, consequently, the drag terms in the currents (\ref{22}),
(\ref{23})) comes only from the second and third terms in Eq.
(\ref{201n}). Consequently, the drag effect considered in this
paper is solely by the excitations. At the mean field level of
approximation (which can be also formulated in terms of the
Gross-Pitaevsky equation) the effect does not appear, while the
coupling between the components is also present at that level of
approximation. We would note that at the  mean field level the
drag effect of another type may emerge. That effect takes place in
the case when one of the species is subjected by an asymmetric
rotating external potential (see, for instance, \cite{loz}, where
such an effect has been studied with reference to the system of
two coupled traps).

\section{Model of Bose-Einstein qubit with external drag force}
\label{sec3}

It is known that Bose systems in the Bose-Einstein condensed state
may demonstrate Josephson phenomenon \cite{leg}. It this paper we
consider the external Josephson effect that takes place in
two-well Bose systems. It was shown in \cite{10} that in such
systems  one can realize the situation, when two states, that
differ in the expectation value of the relative number operator,
can be used as qubit states.

To include the drag force into the play we consider the following
geometry. Let our two-component system is confined in a toroidal
trap and the Bose cloud of the component 1 (the drag component) is
situated inside and overlaps with the Bose clouds of the component
2 (the drive component). Such a situation can be realized if
$|\gamma_{12}|<\min(\gamma_{11},\gamma_{22})$.

\begin{figure}
%\begin{center}
\includegraphics[height=6cm]{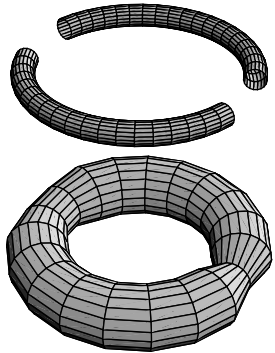}
\caption{\label{fig2}Schematic shapes of Bose clouds for the drag
(top figure) and drive (bottom figure) components. The drag
component is situated inside and overlaps with the drive
component.}
\end{figure}

Deforming the confining potential one can cut the drag component
into two clouds of a half-torus shape (separated by two Josephson
links) leaving the Bose cloud of the drive component uncutted (
Fig. \ref{fig2}). In what follows we use the following notations:
$R_t$ is the large radius of the toroidal trap,  $r_{t1}$ and
$r_{t2}$ are the small radiuses of the toroidal Bose clouds of the
drag and the drive components, correspondingly.

Rotating this trap one can excite a superflow in the drive
component. After the rotation be switched off there will be a
circulating superflow in the drive component and no superflow in
the drag component (at negligible small Josephson coupling).
 The superfluid velocity of the drive component  is
\begin{equation}\label{vor}
  v_2=\frac{N_2 \hbar} {m_2 R_t}
\end{equation}
In (\ref{vor}), we imply that $R_t\gg r_{t1}, r_{t2}$ and neglect,
for simplicity, the effect caused by a dependence of $r_{t2}$ on
the polar angle.

Since $j_1=0$, the phase gradient $\nabla\varphi_1$ should be
nonzero to compensate the drag effect. In the polar coordinates
the $\phi$ component of the phase gradient is given by the
relation
\begin{equation}\label{1004}
  (\nabla \varphi_1)_\phi=-\frac{N_2}{ R_t} f_{dr}=-f_{dr}(\nabla \varphi_2)_\phi,
 \end{equation}
where
\begin{equation}\label{fdr}
  f_{dr}=\frac{m_1}{m_2}\frac{\rho_{dr}}{\rho_{s1}-\rho_{dr}}
\end{equation}
The quantity $f_{dr}$ yields the ratio between the phase gradients
in the drag and the drive components in the situation when the
drag component is in the open circuit (i.e. the current cannot
flow in the circuit). We call this quantity the drag factor.

We imply that $r_{t1}$ and $r_{t2}$ are much larger than the
healing lengths that allows to describe the drag effect in
quasi-uniform approximation. For definiteness, we specify the case
of $\rho_1\ll \rho_2$ and $\rho_2\approx const$ in the overlapping
region. In this case one can neglect the space dependence the drag
factor (see Eqs. (\ref{1002})).

At nonzero Josephson coupling the current $j_1$ can be nonzero,
but it cannot exceed the maximum Josephson current $j_m$. Relation
(\ref{1004}) remains approximately correct at nonzero Josephson
coupling, if an inequality $j_m\ll \hbar\rho_1/(m_1 R_t)$ is
satisfied. Here we specify just such a case. It is important to
emphasize that we consider the situation, when there is only the
external Josephson effect between two half-torus traps, and there
is no internal Josephson effect between the drag and the drive
species.

The drag force  can be considered as an effective vector potential
${\bf A}_{dr}=-\hbar f_{dr} \nabla\varphi_{2}$ (in units of
$e=c=1$) that corresponds to an effective magnetic flux
$\Phi_{dr}=-2\pi \hbar f_{dr} N_2$. Thus, our Bose system is
similar to the Cooper pair box system that implements the
Josephson charge qubit  with the Josephson coupling controlled by
an external magnetic flux \cite{schon}. To extend this analogy we
formulate the model of the Bose-Einstein qubit subjected by the
drag force. In what follows we use the approach of Ref. \cite{10}.

In the two mode approximation the Bose field operators for the
drag component  can be presented in the form:
\begin{equation}\label{1006}
  \hat{\Psi}_1({\bf r},t)=\sum_{l={L},{R}} a_l(t) \Psi_{l}({\bf r}-{\bf
  r}_l),\quad
 \hat{\Psi}_1^+({\bf r},t)=\sum_{l=L,R} a_l^+(t) \Psi_{l}^*({\bf r}-{\bf
  r}_l),
\end{equation}
where $a_{{L}({R})}^+$  and $a_{{L}({R})}$ are the operators of
creation and annihilation of bosons in the condensates confined in
the left(right) half-torus, and $\Psi_L$, $\Psi_R$ are two almost
orthogonal local mode functions $$ \int d^3 r \Psi_{l}^*({\bf r})
\Psi_{l'}({\bf r})\approx\delta_{ll'}, \quad l,l'=L,R$$ that
describe the condensate in the left and right traps \cite{rag}.

Substituting (\ref{1006}) into Hamiltonian (\ref{1}), we obtain
the following expression for the parts of the Hamiltonian that
depends on the operators $a^+_l$ and $a_l$:
\begin{equation}\label{3f}
  H_a=\sum_{l={L},{R}} \left(K_l a_l^+ a_l + \lambda_l a_l^+a_l^+ a_l a_l\right)+
  (J a_{ L}^+ a_{R} + J^* a_{R}^+ a_{L}).
\end{equation}
with
\begin{equation}\label{4f}
  K_l=\int d^3 r  \Psi^*_{l}\left[-\frac{\hbar^2}{2 m} \nabla^2
  +V_{tr}+\gamma_{12} \Psi^*_2\Psi_2\right]\Psi_{l},
\end{equation}
\begin{equation}\label{5f}
  \lambda_l =\frac{\gamma_{11}}{2} \int d^3 r |\Psi_{l}|^4,
\end{equation}
\begin{equation}\label{6f}
  J=\int d^3 r \left[\frac{\hbar^2}{2 m}
  \nabla\Psi_{{L}}^*\nabla\Psi_{{R}}+V_{tr} \Psi_{{L}}^*
  \Psi_{{R}}\right].
\end{equation}

\begin{figure}
%\begin{center}
\includegraphics[height=6cm]{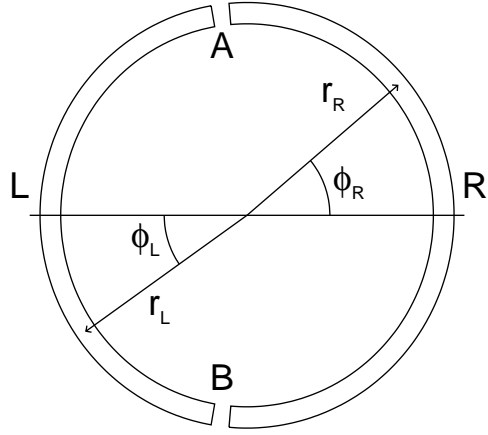}
\caption{\label{fig3}Left (L) and right(R) half-torus of the drag
component, separated by Josephson links A and B.}
\end{figure}

The functions $\Psi_L$ and $\Psi_R$ contain the phase factors
$e^{i\varphi_L(\bf r)}$ and $e^{i\varphi_R(\bf r)}$,  where the
phases satisfy Eq. (\ref{1004}). Taking these factors into
account, one can choose the following basis for the one mode
functions
\begin{equation}\label{1007}
\Psi_{L(R)}({\bf r})=|\Psi_{L(R)}({\bf r})| \exp\left[-i N_2
f_{dr} \phi_{L(R)}({\bf r})\right],
\end{equation}
where $\phi_{L}$,$\phi_{R}$ are the polar angles counted from the
centers of $L$ and $R$ half-torus, correspondingly (see Fig.
\ref{fig3}). The angles $\phi_{L(R)}(\bf r)$, defined as shown in
Fig. \ref{fig3}, satisfy the relation
\begin{equation}\label{fff}
  \phi_{R}({\bf r}_A)-\phi_{L}({\bf r}_A)=\phi_{L}({\bf
r}_B)-\phi_{R}({\bf r}_B)=\pi,
\end{equation}
where ${\bf r}_A$ and ${\bf r}_B$ are the radius-vectors of
Josephson links.

Substituting (\ref{1007}) into Eq. (\ref{6f}), using Eq.
(\ref{fff}) and taking into account that the functions $\Psi_L$
and $\Psi_R$ overlap in a small vicinity of A and B links, we
obtain the following expression for the Josephson coupling
parameter:
\begin{equation}\label{1009a}
  J=(J_A+J_B)\cos\left(\pi
\frac{\Phi_{dr}}{\Phi_0}\right)+i(J_A-J_B)\sin\left(\pi
\frac{\Phi_{dr}}{\Phi_0}\right),
\end{equation}
where $\Phi_0=2\pi\hbar$ is the "flux quantum" and
\begin{equation}\label{1010}
  J_{A(B)}\approx \int_{V_{A(B)}} d^3 r \left[\frac{\hbar^2}{2 m}
  \nabla|\Psi_ L|\nabla|\Psi_R|+V_{tr} |\Psi_L|
  |\Psi_R|\right].
\end{equation}
Here $V_{A}$ and $V_{B}$ are the areas of overlapping of two one
mode functions at links $A$ and $B$, correspondingly.

Considering the Hilbert space in which the total number operator
\begin{equation}\label{1012}
  \hat{N}=a_{ L}^+a_{ L}+a_{ R}^+a_{ R}
\end{equation}
is a conservative quantity ($\hat{N}=N$) we present the
Hamiltonian (\ref{3f}) in the following form
\begin{equation}\label{1013}
  H_a=E_c (\hat{n}_{RL}-n_g)^2+(Ja_{L}^+ a_{R} +
  h.c.) + const ,
\end{equation}
where
\begin{equation}\label{1011}
  \hat{n}_{RL}=\frac{a_{R}^+a_{R}-a_{L}^+a_{L}}{2}
\end{equation}
is the number difference operator,
\begin{equation}\label{1014}
  E_c=\lambda_{R}+\lambda_{L}
\end{equation}
is the interaction energy, and the quantity
\begin{equation}\label{1015}
  n_g=\frac{1}{2 E_c}\left[K_{L}-K_{R}+(N-1)(\lambda_{L}-
  \lambda_{R})\right]
\end{equation}
describes an asymmetry of L and R half-tore.

In what follows we imply that the system is in the Fock regime
\cite{leg} ($|J|N\ll E_c$) and use the number representation
 $$ |n_{RL}\rangle\equiv|n_
R,n_L\rangle \equiv |\frac{N}{2}+n_{RL},\frac{N}{2}-n_{RL}\rangle.
$$ In this representation the first term in (\ref{1013}) is
diagonal. The second term in (\ref{1013}) can be considered as
small nondiagonal correction. But if  $n_g$ is biased  near  one
of degeneracy points
\begin{equation}\label{1016}
  n_{deg}=\cases{M+\frac{1}{2}& for even $N$\cr M& for odd $N$}
\end{equation}
(where $M$ is integer and $|M|<N/2$), the second term in
(\ref{1013}) results in a strong mixing of two lowest states
($|\uparrow\rangle=|n_{deg}+ 1/2\rangle$ and
$|\downarrow\rangle=|n_{deg}-
  1/2\rangle$)
and the low energy dynamics of the system can be described by a
pseudospin Hamiltonian
\begin{equation}\label{1019}
  H_{eff}=-
  \frac{\Omega_x}{2}\hat{\sigma}_x-\frac{\Omega_y}{2}\hat{\sigma}_y-
  \frac{\Omega_z}{2}\hat{\sigma}_z,
\end{equation}
where $\hat{\sigma}_i$ are the Pauli operators, and
\begin{eqnarray}\label{1020a}
 \Omega_x=-(J_A+J_B)\sqrt{(N+1)^2-4 n_{deg}^2}\cos\left(\pi
\frac{\Phi_{dr}}{\Phi_0}\right),\cr
\Omega_y=-(J_A-J_B)\sqrt{(N+1)^2-4 n_{deg}^2}\sin\left(\pi
\frac{\Phi_{dr}}{\Phi_0}\right),\cr \Omega_z=2E_c(n_g-n_{deg})
\end{eqnarray}
are the components of the pseudomagnetic field. In experiments one
can control the parameters $n_g$, $J_A$ and $J_B$ independently
and, consequently, the pseudomagnetic field ${\bm \Omega}(t)$  can
be switched arbitrary. It mean that Eq. (\ref{1019}) represents
the standard Hamiltonian of the qubit system. The parameters of
the qubit (\ref{1019}) depend on the "drag flux" $\Phi_{dr}$.
Therefore, one can determine its value from the measurement of the
state of the system after a controlled evolution of a certain
reproducible initial state.

Let us consider two possibilities. For definiteness, we specify
the case of odd $N$ and the degeneracy point $n_{deg}=0$.

If the Josephson coupling are switched off and $n_g$ is switched
on to some positive value (much less than unity) the system is
relaxed to the state $|\psi_{in}\rangle=|\uparrow\rangle$. This
state can be used as the reproducible initial state. The quantity
should be measured  is the expectation value of the number
difference operator. In the initial state the expectation value of
this operator is $n_{RL}=1/2$

When the system is switched suddenly to the degeneracy  point
$n_g=0$ and the Josephson couplings are switched on for some time
$\tau$ the initial state evolves to another state with another
$n_{RL}$.

If one sets $J_A=J_B=J$ the result of evolution
($|\psi_{f}\rangle=U |\psi_{in}\rangle$) is described by the
unitary operator $$ U_1(\tau)=\left(\matrix{\cos(\alpha_1 \tau)&
-i\sin(\alpha_1 \tau)\cr -i \sin(\alpha_1 \tau)& \cos(\alpha_1
\tau)}\right)$$ where $\alpha_1=(J/\hbar)(N+1)\cos(\pi
\Phi_{dr}/\Phi_{0})$. One can see that at time of evolition
$\tau=\tau_1=\pi/(4|\alpha_1|)$ the expectation value of the
number difference operation will be equal to zero.

For the case $J_A=J$ and $J_B=0$ the operator of evolution reads
as $$ U_2(\tau)=\left(\matrix{\cos(\alpha_2 \tau)& -i
e^{-i\pi\Phi_{dr}/\Phi_{0}} \sin(\alpha_2 \tau)\cr -i
e^{i\pi\Phi_{dr}/\Phi_{0}}\sin(\alpha_2 \tau)& \cos(\alpha_2
\tau)}\right)$$ with $\alpha_2=(J/2\hbar) (N+1)$. Respectively,
the expectation value ${n}_{RL}$ will be equal to zero at
$\tau=\tau_2=\pi/(4\alpha_2)$

The ratio $\tau_2/\tau_1=|\cos(\pi \Phi_{dr}/\Phi_{0})|/2$ depends
only on $\Phi_{dr}$ and the quantity $\Phi_{dr}$ can be extracted
from the measurements of $\tau_1$ and $\tau_2$.   It is important
to note that to provide this scheme one should control only the
ratio of $J_A$ and $J_B$, but not their absolute values.

Another possibility can be based on detection of the Berry phase
\cite{ber}. Eq. (\ref{1019}) contains all three components of the
field ${\bm \Omega}$ and they can be controlled independently. The
general scheme of detection of the Berry phase in such a situation
was proposed  \cite{8}. A concrete realization of this scheme in
the Josephson charge qubit was described in  \cite{9}. Here we
extend the ideas of \cite{8,9} to the case of the "dragged"
Bose-Einstein qubit.

We start from the same initial state and switch to $J_A=J_B=J$ and
$n_g=0$. The initial state $|\uparrow\rangle$ can be presented as
the superposition of two instantaneous eigenstates
$|e_a\rangle=(|\uparrow\rangle+|\downarrow\rangle)/\sqrt{2}$ and
$|e_b\rangle=(|\uparrow\rangle -|\downarrow\rangle)/\sqrt{2}$:
\begin{equation}\label{301}
|\psi_{in}\rangle=\frac{1}{\sqrt{2}}(|e_a\rangle+|e_b\rangle).
\end{equation}
An adiabatic cyclic evolution of the parameters of the Hamiltonian
(\ref{1019})  results in appearance of the Berry phase in the
$|e_a\rangle$ and $|e_b\rangle$ eigenstates, if the vector ${\bm
\Omega}$ subtends a nonzero solid angle at the origin.

Let us consider the following 4 stage cyclic adiabatic evolution
starting from the point $J_A=J_B=J$ and $n_g=0$: 1 - $J_B$ is
switched off; 2 - $J_A$ is switched off and simultaneously $n_{g}$
is switched to $n_{g1}>0$; 3 - $n_g$ is returned to the same
degeneracy point ($n_g=0$) and $J_B$ is switched to $J_B=J$; 4 -
$J_A$ is switched to $J_A=J$ (all switches should be done slowly:
$\hbar|d{\bf \Omega}/dt|\ll \Omega^2$).

After such an evolution the system arrives at the state
\begin{equation}\label{302}
  |\psi_{m}\rangle=\frac{1}{\sqrt{2}}\left(e^{i\delta_a+i\gamma}|e_a\rangle+
e^{i\delta_b-i\gamma}|e_b\rangle\right),
\end{equation}
where $\gamma=\pi\Phi_{dr}/\Phi_0$ is the Berry phase (equals to
half of the solid angle subtended by ${\bm \Omega}$) and
$\delta_a$, $\delta_b$ are the dynamical phases.

 Elimination of the dynamical phases can performed
 by swapping the eigenstates ($\pi$-transformation) and
repeating the same cycle of evolution in a reverse direction (see
\cite{8}).

 The
$\pi$-transformation can be done by fast switching off the
Josephson coupling and switching on $n_g= n_{g2}>0$ during the
time interval $t_\pi=\hbar\pi/(2 E_c n_{g2})$. After the
$\pi$-transformation the state becomes
\begin{equation}\label{303}
  |\psi_{m\pi}\rangle=-\frac{i}{\sqrt{2}}(e^{i\delta_a+i\gamma}|e_b\rangle+
e^{i\delta_b-i\gamma}|e_a\rangle).
\end{equation}

After the cyclic evolution in the reverse direction we arrive to
the state
\begin{equation}\label{304}
  |\psi_{f}\rangle=-\frac{i}{\sqrt{2}}e^{i(\delta_a+\delta_b)}(e^{2i\gamma}|e_b\rangle+
e^{-2i\gamma}|e_a\rangle).
\end{equation}

One can see that the expectation value of the number difference
operator in the final state (\ref{304})
$n_{RL}=\cos(4\gamma)/2=\cos(4\pi\Phi_{dr}/\Phi_0)/2$ depends only
on $\Phi_{dr}$ and the measurement of this difference allows the
determine the value of the "drag flux".

Thus, the measurements of relative number of atoms in left and
right condensates under controlled evolution of the state of the
system allows  to observe the non-dissipative drag and determine
the drag factor (if the vorticity of the drive component is
known).

\section{Conclusions}
\label{sec5}

We have investigated the  non-dissipative drag effect in
three-dimensional weakly interacting two-component superfluid Bose
gases. The expression for the drag current is derived
microscopically for the general case of two species of different
densities, different masses and different interaction parameters.
It is shown that the  drag current is proportional to the square
root of the gas parameter. The drag effect is maximal at zero
temperatures and it decreases when the temperature increases, but
at temperatures of order of the interaction energy the drag
current remains of the same order as at zero temperature.

We have considered the toroidal double-well geometry, where the
non-dissipative drag influences significantly on the Josephson
coupling between the wells. In the system considered the drag
force can be interpreted as an effective vector potential applied
to the drag component. The effective vector potential is equal to
${\bf A}_{dr}=-\hbar f_{dr} \nabla\varphi_{drv}$ (in units of
$e=c=1$), where $\varphi_{drv}$ is the phase of the drive
component, and $f_{dr}$ is the drag factor. In the toroidal
geometry the effective vector potential can be associated with an
effective flux of external field $\Phi_{dr}=2\pi \hbar f_{dr}
N_v$, where $N_v$ is the vorticity of the drive component. In the
Fock regime the system can be considered as a Bose-Einstein
counterpart of the Josephson charge qubit in an external magnetic
field. The measurement of the state of such a qubit allows to
observe the drag effect and determine the drag factor.

\section*{Acknowledgements}
This work is supported by the INTAS grant No 01-2344.

%\section*{References}

\end{document}